\documentclass[10pt,conference]{IEEEtran}
\ifCLASSINFOpdf
   \usepackage[pdftex]{graphicx}
\else
\fi
\usepackage{graphicx}

\usepackage[cmex10]{amsmath}
\usepackage{array}

\usepackage{amsmath}
\usepackage{amsfonts}
\usepackage{amssymb}
\begin{document}
%
\title{Capacity and Error Rate Analysis of MIMO Satellite Communication Systems in Fading Scenarios}

\author{\IEEEauthorblockN{Ramoni O. Adeogun\\}
\IEEEauthorblockA{School of Engineering and Computer Science\\
Victoria University of Wellington\\
Wellington, New Zealand\\
Email: ramoni.o.adeogun@ieee.org}}


%


\maketitle

\begin{abstract}
In this paper, we investigated the capacity and bit error rate (BER) performance of Multiple Input Multiple Output (MIMO) satellite systems with single and multiple dual polarized satellites in geostationary orbit and a mobile ground receiving station with multiple antennas. We evaluated the effects of both system parameters such as number of satellites, number of receive antennas, and SNR and environmental factors including atmospheric signal attenuations and signal phase disturbances on the overall system performance using both analytical and spatial models for MIMO satellite systems.
\end{abstract}

%
\IEEEpeerreviewmaketitle

\section{Introduction}
Multiple-Input Multiple-Output (MIMO) wireless communications systems have been a focus of academic and industrial research in the last decade due to their potentially higher data rates in comparison with Single-Input Single-Output (SISO) systems \cite{Driessen}. Theoretically, the overall channel capacity can be increased linearly with the number of transmit and receive antennas by using spatial multiplexing schemes \cite{Driessen}. Current focus on satellite communication (SatCom) systems recognizes a demand for higher data rates. Hence, it appears to be appropriate to apply MIMO to SatCom systems in order to increase the available data rate and bandwidth efficiency.

The quality of service (QoS) and data rates requirements of satellite communication systems is recently on the increase. Hence, the application of multiple input multiple output techniques to satellite communication systems appear to be appropriate in order to achieve increased spectral and bandwidth efficiency \cite{Schwarz}. Spatial multiplexing and diversity maximization schemes can be deployed to achieve better spectral efficiencies and bit error rates (BER) when compared to the classical single satellite single receive station systems.

In \cite{Schwarz}, MIMO satellite uplinks and downlinks channel that are optimal in terms of achievable data rates were analyzed. The authors showed that capacity optimization is generally possible for regenerative payload designs using Line of Sight (LOS) channel models. These analysis were extended to a number of MIMO satellite communication systems in \cite{Schwarz2} and the scope was further extended to general  case of satellites with transparent communication payloads component. A cluster based channel model was proposed for MIMO satellite formation systems in \cite{Ramoni2013}. Based on the standardized models for terrestrial multiple input multiple output (MIMO) systems, the authors proposed a spatial model and analysed the capacity of formation flying satellite systems.

In this contribution, we analyse the performance of satellite communication systems with multiple cooperating satellites in geostationary orbit (GEO) and single or multiple antennas at the ground receiving station. The analysis in this paper is based on three different modelling approaches for land mobile satellite systems.

The remaining part of this paper is organized as follows. In Section II, we present the system model for MIMO satellite systems. A review of the propagation channel models considered in the paper is presented in section III. In Section IV, we derive expressions for channel capacity and bit error rates with MPSK modulation scheme. Simulation results and discussions are presented in section V. Finally, we draw conclusion in Section VI.

\section{System Model}
In this section, we present the system model for single satellite, multiple receive antenna systems (SS-MRA) and multiple satellite multiple receive antenna systems (MS-MRA).
\subsection{Single Satellite - Multiple Receive Antennas (SS-MRA)}
Consider the downlink of a Land-mobile satellite receive diversity system consisting of a single dual polarized satellite antenna and a mobile receive station with $M$ non-polarized antennas. The channel impulse response between the satellite and the mobile receive station can be modelled as an $M\times 2$ MIMO communication channel
\begin{equation}
\label{eq1}
\mathbf{H}=\begin{bmatrix}
h_{11} & h_{12}\\
h_{21} & h_{22}\\
\vdots & \vdots\\
h_{M1} & h_{M2}
\end{bmatrix}
\end{equation}
where $h_{ij}$ is the channel between the $j$-th transmit polarization and the $i$-th receive antenna. The received signal at the mobile receive antennas is given by
\begin{equation}
\label{eq2}
\begin{bmatrix}
y_1\\
y_2\\
\vdots\\
y_M
\end{bmatrix}
=\begin{bmatrix}
h_{11} & h_{12}\\
h_{21} & h_{22}\\
\vdots & \vdots\\
h_{M1} & h_{M2}
\end{bmatrix}\begin{bmatrix}
x_1\\
x_2
\end{bmatrix}+\begin{bmatrix}
n_1\\
n_2\\
\vdots\\
n_M
\end{bmatrix}
\end{equation}
A matrix representation for the receive signal model in \eqref{eq2} is thus
\begin{equation}
\mathbf{y}=\mathbf{H}\mathbf{x}+\mathbf{n}
\end{equation}
where $\mathbf{y}=[y_1,y_2,\cdots,y_M]^T$ is an $M\times 1$ vector of the received signals at the $M$ receive antennas, $\mathbf{x}=[x_1,x_2]^T$ is a vector of transmitted symbols on the two polarizations of the satellite antenna and $\mathbf{n}=[n_1,n_2,\cdots,n_M]^T$ is an $M\times 1$ noise vector assumed to be complex Gaussian random variables with zero mean and variance $\sigma^2$.
\subsection{Multiple Satellite - Multiple Receive Antennas (MS-MRA)}
We consider a satellite diversity system comprising of $N$ dual polarized satellites and a mobile ground receiving station with $M$ equally spaced antennas. This corresponds to a $2N\times M$ multiantenna wireless system. However, since the satellites antennas are not co-located, the relative delay between signal transmission from each satellites need to be accounted for in the system model \cite{Schwarz2}. The received signal at the mobile station can therefore be modelled as
\begin{equation}
\mathbf{y}(t)=\left[\mathbf{H}_{s1}(t)\,\, \mathbf{H}_{s2}(t)\,\, \cdots \,\, \mathbf{H}_{sN}(t)\right]\begin{bmatrix}
\mathbf{x}^1(t)\\
\mathbf{x}^2(t-\tau_2)\\
\vdots\\
\mathbf{x}^N(t-\tau_N)
\end{bmatrix}+\mathbf{n}(t)
\end{equation}
\begin{figure}
\begin{center}
\label{fig:MIMOsat}
\includegraphics[width = 1\columnwidth]{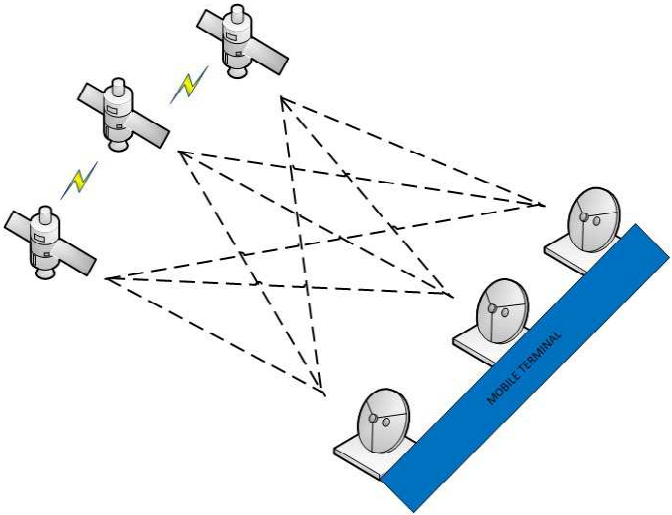}
\caption{Multiple Satellites Multiple Antennas Ground Receive Station System}
\end{center}
\end{figure}
where $\mathbf{H}_{si}$ is the $2\times M$ impulse response matrix for the channel between the $i$-th satellite and the $M$ receive antennas, $\mathbf{y}(t)=\left[y_1(t),y_2(t),\cdots,y_M(t)\right]^T$ are the received signals, $\mathbf{x}^i(t)=[x_1^i(t) \,\, x_2^i(t)]^T$ are the transmitted signals on the two polarizations of satellite $i$ and $\tau_n$ is the relative delay experienced by signals from the $n$th satellite with respect to the reference satellite.

\section{Channel Models}
We consider three different models for our evaluations in this paper. The models are the cluster based spatial satellite MIMO model \cite{Ramoni2013}, Loo distribution based analytical model \cite{Loo85,LooB98} and the physical - statistical land mobile satellite model \cite{Schwarz}. A brief description of the satellite channel models is presented in this section.
\subsection{Cluster Based MIMO Satellite Model}
In \cite{Ramoni2013}, a cluster based MIMO model was proposed for MIMO satellite systems using the concept of clustering\footnote{A cluster is generally considered as a group of propagation paths sharing common angle of arrivals and/or delays of arrival. In the cluster based approach for satellite models, it is assumed that paths within a cluster share closely spaced delays of arrival.} in the standardized WINNER II/3GPP model for terrestial MIMO systems. The spatial model is given by \cite{Ramoni2013}
\begin{equation}
\label{eq:model1}
h_{nm}(t)=\sqrt{\frac{K}{K+1}}h_{nm}^{LOS}(t)+\sqrt{\frac{1}{K+1}}\sum_{p=2}^{P}g^{(p)}_{nm}(t)\delta(\tau-\tau_p)
\end{equation}
where $K$ is the Ricean K-factor, $h_{nm}^{LOS}$ is the line of sight (LOS) component of the channel impulse response between the nth satellite and the mth ground receiver antenna. The second term in the RHS of \eqref{eq:model1} is the non-line-of-sight (NLOS) component of the channel which is modelled as a summation of P clusters, each cluster comprising of R rays. The LOS and NLOS component are modelled as
\begin{align}
h_{nm}^{LOS}(t)=\sqrt{P_p\exp(j\Phi_{p}).G_R(\theta_{p}).\sigma_{p}.\mathcal{P}_{p}.G_T(\phi_{p})}.\nonumber\\
\exp{(\frac{j2\pi}{\lambda}(d_s\sin(\theta_{p})+d_m\sin(\phi_{p}+\Upsilon_{p})))}.\nonumber\\
\exp{(\frac{-j2\pi V_m}{\lambda}\cos(\vartheta_v-\theta_{p}))}
\end{align}
and
\begin{align}
g^{(p)}_{nm}(t)=\sqrt{\frac{P_p}{R}}\sum_{r=1}^{R}\sqrt{\exp(j\Phi_{rp}).G_R(\theta_{rp}).\sigma_{rp}.\mathcal{P}_{rp}.G_T(\phi_{rp})}.\nonumber\\
\exp{(\frac{j2\pi}{\lambda}(d_s\sin(\theta_{rp})+d_m\sin(\phi_{rp}+\Upsilon_{rp})))}.\nonumber\\
\exp{(\frac{-j2\pi V_m}{\lambda}\cos(\vartheta_v-\theta_{rp}))}
\end{align}
$P_p$ is the normalised power of the p-th multipath component(MPC), $R$ is the number of rays within each cluster(assumed constant in the model), $\Phi$ is the ionospheric power loss compensation factor for each ray in the clusters, $G_R(\theta)$ is the ground receive station array gain for each antenna in the array, $\theta_{rp}$ is the AOA of the rth ray in the pth cluster, $\sigma$ is the shadow fading coefficient of the rays, $\mathcal{P}$ is the path loss, $G_T(\phi)$ is the satellite transmit antenna response for rays with AOD $\phi$, $\lambda$ is the wavelength,  $d_s$ is the inter-satellite spacing, $\theta_{rp})$ is the AOD of the rth ray of the pth cluster,$d_m$ is the spacing between the antennas on the mobile ground receiving station antenna array, $\phi_{rp}$ is the AOA of the rth ray in the pth cluster,$V_m$ is the velocity of the receive station, $\Upsilon$ is the ionospheric angular deviation compensation and $\vartheta$ is the direction of motion of the ground receive station.
\subsection{Free Space LOS Model}
The free space MIMO satellite model consider the line of sight (LOS) component of the fading channel. Each entry of the MIMO impulse response matrix is defined by \cite{Schwarz}
\begin{equation}
\label{eq3}
\mathbf{H}_{ij}=\alpha_{ij}\exp(-jk_0f_cr_{ij})
\end{equation}
where $f_c$ is the carrier frequency, $r_{ij}$ is the geometrical distance between the $j$-th satellite transmit antenna and $i$-th mobile ground receive station antenna, $k_0=\frac{2\pi}{v_0}$ is the wave number, $v_0$ is the free space speed of light and $\alpha_{ij}$ is the complex attenuation of the propagation path defined as
\begin{equation}
\label{eq4}
\alpha_{ij}=\frac{1}{2k_of_cr_{ij}}\exp(j\phi)
\end{equation}
where $\phi$ is the phase of the carrier assumed equal for all antenna pairs. Since the approximation $r_{ij}\approx r\pm 3km \forall i,j$ is applicable to the satellite systems considered in this paper, the channel path gains can therefore be approximated by \cite{Schwarz10}
\begin{equation}
\label{eq4a}
|\alpha_{ij}|\approx|\alpha| = C; \forall i,j
\end{equation}
where $C$ is a constant and $|a|$ denotes the absolute value of $a$.
\subsection{Analytical MIMO Satellite Model}
The Loo distribution \cite{Loo85} is often used for the analytical modelling of land mobile satellite channels. The MIMO impulse for the multi-polarization and multiantenna channel considered in this paper can therefore be modelled as a summation of two parts
\begin{align}
\label{eq5}
\mathbf{H}
&=\begin{bmatrix}
\tilde{h}_{11} & \tilde{h}_{12}\\
\tilde{h}_{21} & \tilde{h}_{22}\\
\vdots & \vdots\\
\tilde{h}_{M1} & \tilde{h}_{M2}
\end{bmatrix}+
\begin{bmatrix}
\overline{h}_{11} & \overline{h}_{12}\\
\overline{h}_{21} & \overline{h}_{22}\\
\vdots & \vdots\\
\overline{h}_{M1} & \overline{h}_{M2}
\end{bmatrix}\nonumber\\
&=\tilde{\mathbf{H}}+\overline{\mathbf{H}}
\end{align}
where $\tilde{\mathbf{H}}$ models the shadowing effect of the channel and its entries are generated using the Log-normal distribution and $\overline{\mathbf{H}}$ is the multipath component of the channel with Rayleigh distributed entries. The Loo distribution based analytical models characterize the channel statistics using probability density function (pdf) and cumulative distribution function (CDF). A general assumption is that the propagating wave undergo both attenuation and scattering/reflection. As given in \eqref{eq5}, the complex channel envelope is a summation of Rayleigh and log-normal faded envelopes. The pdf of the channel is defined as \cite{Loo85}
\begin{equation}
 f(r) = \left\{
  \begin{array}{l l}
    \frac{1}{r\sqrt{2\pi\sigma_r^2}}\exp\left[-\frac{(\log r -\mu)^2}{2\sigma^2_r}\right] & \quad \text{for $r>>c_o$}\\
    \frac{r}{c_o}\exp\left[-\frac{r^2}{2c_o}\right] & \quad \text{for $r<<c_o$}
  \end{array} \right.
\end{equation}
where $\mu$ and $\sigma_r^2$ are the mean and variance of the received signal envelope, respectively. $c_o$ gives the average power of the scattered component of transmitted signal.
\section{Channel capacity and BER}
In this section, we present the channel capacity and theoretical bit error rate (BER) expressions.
\subsection{Channel Capacity}
The channel capacity for a narrowband MIMO system without channel state information at the transmitter (CSIT) is generally given by Telatar's spectral efficiency equation \cite{Telatar}
\begin{equation}
\label{cap1}
C=\log_2\left[\det(\mathbf{I}_{M\times M} +\rho\mathbf{H}\mathbf{H}^H)\right]
\end{equation}
where $(.)^H$ denotes the Hermittan transpose of a matrix and $\rho$ is the linear signal-to-noise ration value computed from the logarithmic SNR by
\begin{equation}
\label{cap2}
\rho = 10^{(\frac{SNR}{10})}
\end{equation}
Similar to \cite{Schwarz}, $\rho$ is defined as the ratio of the transmit power at each of the satellite antenna and the noise power at each mobile ground receive antenna. The decibel value of the SNR in \eqref{cap2} is defined as
\begin{equation}
\label{cap3}
SNR = EIRP + G_T-\mathcal{K}-\mathcal{B}
\end{equation}
where EIRP is the effective isotropic radiated power, $G_T$ is the satellite figure of merit, $\mathbf{K}$ is the dB equivalent of Boltzmann's constant and $\mathcal{B}$ is the downlink transmission bandwidth.
\subsection{Bit Error Rate (BER)}
Following the analysis and derivations in \cite{JinChua}, a closed form approximation for the probability of error for MPSK modulated transmission in additive white Gaussian noise (AWGN) is given as \cite{JinChua}
\begin{align}
\label{BER1}
P_{ERR}&=\gamma\sum_{k=1}^{\min(2,[M/4])}Q\left(\sqrt{2\sigma x}\sin\left(\frac{(2k-1)\pi}{M}\right)\right)\\
\gamma&=\frac{2}{\max(\log_2 M,2)}
\end{align}
where $M$ is the constellation size, $\sigma$ is the SNR per symbol, $x$ is a chi-square distributed random variable and $[M/4]$ denotes the smallest integer greater than or equal to $M/4$.
Assuming that the mobile ground receive station uses a zero forcing (ZF) receiver, the MPSK BER can be obtained by integrating the error probability in \eqref{BER1} over $x$
\begin{equation}
\label{BER2}
MPSK_{BER}=\int_{0}^{\infty}P_{ERR}P_X(x)\,dx
\end{equation}
where $P_X(x)$ is the chi-square probability distribution function. It can be shown that a closed form expression for \eqref{BER2} is \cite{Cheng}
\begin{align}
\label{BER3}
MPSK_{BER}&=\frac{2}{\max(\log_2 M,2)}\sum_{k=1}^{\min(2,[M/4])}\left[\frac{1}{2}(1-\mu_k)\right]^{U}\nonumber\\
&.\sum_{\ell=1}^{U-1}\begin{pmatrix}
U-1+\ell\\
\ell
\end{pmatrix}\left[\frac{1}{2}(1+\mu_k)\right]^\ell
\end{align}
where $U=N-M+1$ and $\mu_k$ is given by
\begin{equation}
\label{BER4}
\mu_k = \sqrt{\frac{\sin^2((2k-1)\pi/M)\sigma}{1+\sin^2((2k-1)\pi/M)\sigma}}
\end{equation}
\section{Simulation Results}
\begin{figure}[htb]
\centering
\includegraphics[width = 0.9\columnwidth]{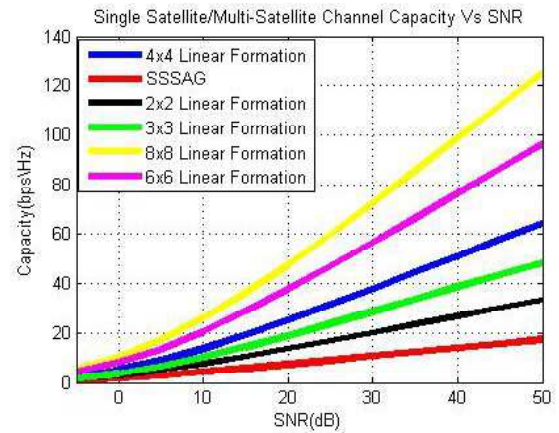}
\caption{MIMO Satellite Capacity Versus Signal to Noise Ratio (SNR) Using the Cluster Based Satellite Channel Model: SSSAG denotes Single Satellite Single Antenna Ground Receive Station}
\label{Capacity1}
\end{figure}

\begin{figure}
\centering
\includegraphics[width = 0.9\columnwidth]{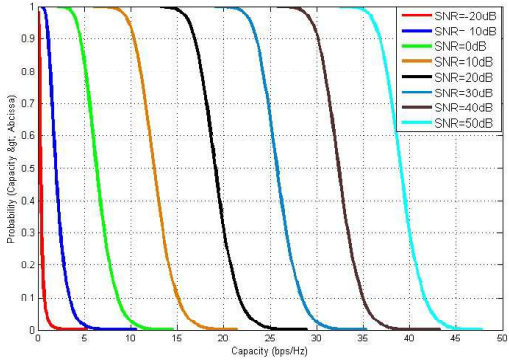}
\caption{Complimentary Capacity Cummulative Distribution Function for a Single Dual Polarized Satellite and Ground Receive Station with four Antennas ($2\times 4$ MIMO) Using the Loo Distribution based analytical MIMO Satellite Model at different signal to noise ratio (SNR)}
\label{Capacity2}
\end{figure}

\begin{figure}
\centering
\includegraphics[width = 0.9\columnwidth]{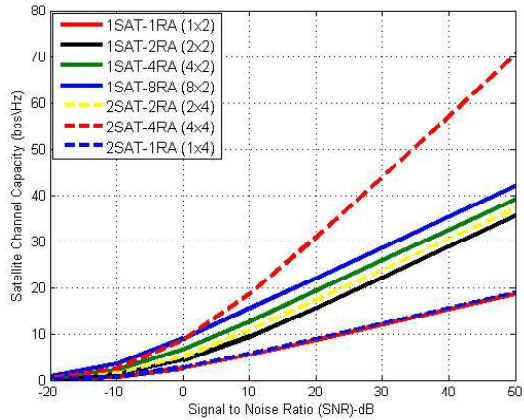}
\caption{MIMO Satellite Capacity Versus SNR for Single Satellite Multiple Receive Antenna Ground Station (SS-MRA) and Multiple Satellites Multiple Ground Receive Antennas Ground Station (MS-MRA) Using the Loo-Distribution Based Analytical Satellite Model.}
\label{Capacity3}
\end{figure}

\begin{figure}
\centering
\includegraphics[width = 0.9\columnwidth]{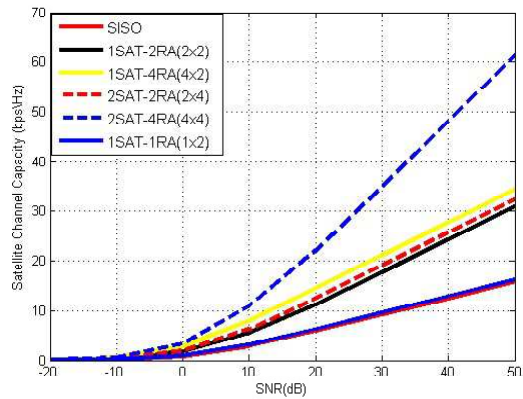}
\caption{MIMO Satellite Capacity Versus SNR for Single Satellite Multiple Receive Antenna Ground Station (SS-MRA) and Multiple Satellites Multiple Ground Receive Antennas Ground Station (MS-MRA) Using the Line of Sight (LOS) Satellite Model.}
\label{Capacity4}
\end{figure}

\begin{figure}
\centering
\includegraphics[width = 0.9\columnwidth]{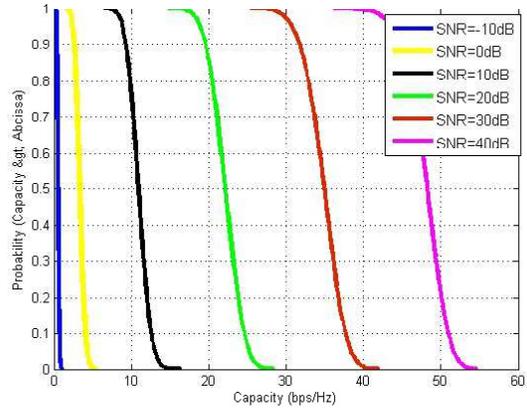}
\caption{Complimentary Capacity Cummulative Distribution Function for a Single Dual Polarized Satellite and Ground Receive Station with four Antennas ($2\times 4$ MIMO) Using the Line of Sight (LOS) MIMO Satellite Model at different SNR for a 2 Dual Polarized Satellites - 4 Receive Antennas (4x4) System}
\label{Capacity5}
\end{figure}

In this section we present simulation results for the capacity and BER of different configurations of MIMO satellite systems with the models present in Section III. The simulation parameters for the simulations are shown in Table \ref{tab:SimulationPar} except where otherwise stated.
\begin{table}[!t]
\centering
\caption{Simulation Parameters}
\begin{tabular}{|c|c|}
\hline
Parameters & Value \\
\hline
\hline
Satellite Orbit & Geostationary \\
\hline
Satellite Location & $13^o$ E \\
\hline
Intersatellite Spacing & 6m \\
\hline
Carrier frequency & 14GHz \\
\hline
Receive antenna spacing (2 satellites)  & 68.2km\\
\hline
Ground station antenna location & $11.1^o E$, $47.8^o N$ \\
\hline
Modulation & BPSK, QPSK with gray mapping \\
\hline
Channel Models & See Section III \\
\hline
Environment & Typical Urban \\
\hline
\end{tabular}
\label{tab:SimulationPar}
\end{table}
The intersatellite spacing for systems with $M>2$ receive antennas is calculated using the equation \cite{Schwarz}\footnote{Detailed derivations and justification can be found in \cite{Schwarz}}
\begin{equation}
\label{sim1}
d_s^{M\times 2}=d_s^{2\times 2}\times \frac{2}{M}
\end{equation}
In Figure~\ref{Capacity1}, we present the capacity (in bps/Hz) as a function of SNR for linear formation multiple satellite system using the cluster based spatial channel model. The number of satellites and receive antenna elements is varied between 1 and 8. As shown in the figure, increasing the signal to noise ratio (SNR) increases the channel capacity for all antenna sizes as expected. The capacity also increases with increase in the number of satellites and/or receive station antenna elements. For instance, the capacity difference between a $2\times 2$ and $4\times 4$ satellite system at $SNR = 30\,dB$ is about 10\,dB. Figure \ref{Capacity2} present the complementary capacity cumulative distribution function (CCDF) for a dual polarized satellite system and a mobile ground receive station with four antenna elements (corresponding to a $2\times 4$ MIMO system) at different signal to noise ratio (SNR) levels. The CDF plots show that the variance of the channel capacity is considerably small for each SNR level. The capacity increase with SNR can also be clearly observed from Fig. \ref{Capacity2}. In figure \ref{Capacity3}, we compare the capacity for different number of satellites and receive antennas using the Loo-distribution based analytical satellite channel model for single and multi-satellite scenarios. Clearly, the channel capacity also shows an increasing trend with both increase in SNR and antenna sizes. We present a plot of the MIMO satellite channel capacity versus SNR for both single satellite multiple receive antenna ground station (SS-MRA) and multiple satellites multiple receive antenna ground station (MS-MRA) using the line of sight (LOS) approximation model in figure~\ref{Capacity4}. As can be observed from the figure, the channel capacity obtained using the LOS approximation model shows a similar trend and compare well with the capacity for similar scenarios using the cluster based and analytical channel models. In figure~\ref{Capacity5} present the complementary capacity cumulative distribution function (CCDF) for a dual polarized satellite system and a mobile ground receive station with four antenna elements (corresponding to a $2\times 4$ MIMO system) at different signal to noise ratio (SNR) levels using the line of sight (LOS) approximation model. Finally, we plot the bit error rate (BER) versus signal to noise ratio (SNR) for a two-satellite two receive antenna system using the three types of model described in section III. As shown in the figure, the cluster based model gives lower BER at higher SNR. However, no significant difference is observed between the BER curves for the three channel models at low SNR region. Summarily, the results presented in this section shows that the spectral efficiency of satellite systems can be significantly improved by having multiple satellites and multiple antennas at the ground station.
\begin{figure}
\centering
\includegraphics[height=2.5in,width=1\columnwidth]{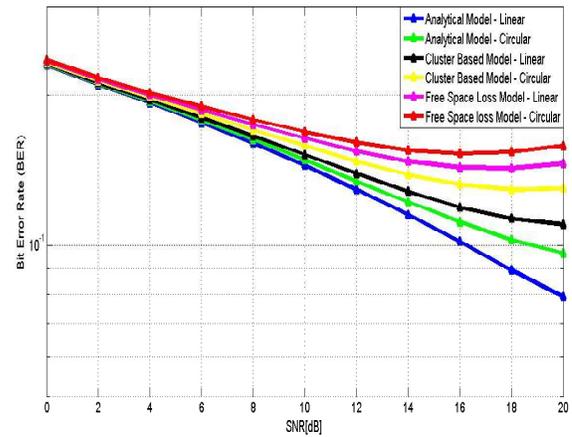}
\caption{Bit Error Rate (BER) versus Signal to Noise Ratio (SNR) in dB for a Two-Satellite Two- Receive Antenna System using Cluster Based Model, Free Space Loss Model and Analytical MIMO Satellite Model.}
\label{fig:figBER}
\end{figure}
\section{Conclusion}
Multiple input multiple output dual polarized satellite systems can provide increased spectral efficiency and improved bit error rate (BER) compared to the classical single satellite systems. In this paper, we analyzed the capacity and BER of different multiple satellite scenarios using different models. Simulation results showed that increasing the number of satellite and/or ground receive station antennas can significantly increase the capacity and decrease the bit error rate.




\bibliographystyle{IEEEtran}
%

\end{document}